\documentclass[5p,authoryear,twocolumn]{elsarticle}

\usepackage{amsmath}
\usepackage{algorithm}
\usepackage{algorithmic}
\usepackage{graphicx}

\begin{document}

\title{Real-Time Dedispersion for Fast Radio Transient Surveys, using Auto Tuning on Many-Core Accelerators}

\author[vu,astron]{Alessio Sclocco}
\ead{a.sclocco@vu.nl}
\author[astron,api]{Joeri van Leeuwen}
\ead{leeuwen@astron.nl}
\author[vu]{Henri E. Bal}
\ead{h.e.bal@vu.nl}
\author[nlesc]{Rob V. van Nieuwpoort}
\ead{r.vannieuwpoort@esciencecenter.nl}

\address[vu]{Faculty of Sciences, Vrije Universiteit Amsterdam, Amsterdam, the Netherlands}
\address[astron]{ASTRON, the Netherlands Institute for Radio Astronomy, Dwingeloo, the Netherlands}
\address[api]{Astronomical Institute ``Anton Pannekoek'', University of Amsterdam, Amsterdam, the Netherlands}
\address[nlesc]{NLeSC, Netherlands eScience Center, Amsterdam, the Netherlands}

\begin{abstract}
Dedispersion, the removal of deleterious smearing of
  impulsive signals by the interstellar matter, is one of the most
  intensive processing steps in any radio survey for pulsars and fast
  transients.  We here present a study of the parallelization of this
  algorithm on many-core accelerators, including GPUs
  from AMD and NVIDIA, and the Intel Xeon Phi. We find that
  dedispersion is inherently memory-bound. Even in a perfect scenario,
  hardware limitations keep the arithmetic intensity low, thus
  limiting performance. We next exploit auto-tuning to adapt
  dedispersion to different accelerators, observations, and even
  telescopes. We demonstrate that the optimal settings differ between
  observational setups, and that auto-tuning significantly improves
  performance. This impacts time-domain surveys from Apertif to SKA\@.
\end{abstract}


\maketitle

\section{Introduction}
\label{sec:introduction}

Astronomical phenomena such as pulsars \citep{hbp+68} and fast radio
bursts \citep[FRBs; ][]{lbm+07} consist
of millisecond-duration impulsive signals over a broad radio-frequency
range.
As the electromagnetic waves propagate through the interstellar
material (ISM) between us and the source, they are dispersed \citep{lorimer2005}.
This causes lower radio frequencies to arrive progressively later, and without correction this results in a loss of signal-to-noise, that often makes the source undetectable when integrating over a wide observing bandwidth.
This frequency-dependent delay can be removed in a process called
\textit{dedispersion}. Complete removal can be achieved by reverting
all phases through a convolution of the signal with the inverse of the modeled
ISM \citep[coherent dedispersion; ][]{hr75}. Incomplete but much
faster removal, especially when many dispersion measure trials are
required, can be achieved by appropriately shifting in time the signal
frequency channels (incoherent dedispersion; from now on referred to plainly as dedispersion).
This dedispersion is a basic algorithm in high-time-resolution radio
astronomy, and one of the building blocks of surveys for fast
phenomena with modern radio telescopes
such as the Low Frequency Array \citep[LOFAR; ][]{ls10, sha+11} and the
Square Kilometer Array \citep[SKA; ][]{carilli2004}.
In these surveys, the dispersion measures are a priori unknown, and
can only be determined in a brute-force search.
This search generally runs on off-site
supercomputers. These range from e.g., the CM-200 in the \citet{fcwa95}
Arecibo survey, to
gSTAR for the Parkes HTRU \citep{kjs+10}, and Cartesius for
the LOFAR LOTAAS \citep{clh+14} surveys. In the latter the
dedispersion step amounts to over half of
all required pulsar-search processing. For the SKA, this processing
will amount to many PFLOPS for both SKA-Mid \citep[cf.\,][]{magr14} and
  SKA-Low \citep{kbk+14}.

Above and beyond these pure compute requirements, the similar and
often simultaneous search for FRBs demands that this dedispersion is
done near real time. Only then can these fleeting events be immediately
followed up by telescopes at other energies \citep[][]{pbb+15}.

We aim to achieve the required performance by parallelizing this
algorithm for many-core accelerators.  Compared to similar attempts
made by~\cite{barsdell2012b} and~\cite{armour2012}, we
present a performance analysis that is more complete, and
introduce a novel many-core algorithm that can be tuned for different platforms and
observational setups.  To our knowledge, this is the first attempt at
designing a brute-force dedispersion algorithm that is highly portable and not
fine-tuned for a specific platform or telescope.


To summarize our contributions, in this paper we: (1) provide an
in-depth analysis of the arithmetic intensity (AI) of brute-force dedispersion,
finding  analytically and  empirically that it is
memory bound; (2) show that auto-tuning can
adapt the algorithm to different platforms, telescopes, and
observational setups; (3) demonstrate that
many-core accelerators can achieve real-time performance; (4) quantify the
impact that auto-tuning has on performance; (5) compare different
platforms using a real-world scientific application; and (6) compare
the performance of OpenCL and OpenMP for the Intel Xeon devices.

In Section~\ref{sec:dedispersion_kernel} we describe the
brute-force dedispersion algorithm, our parallel implementation and
its optimizations; and the theoretical analysis of
the dedispersion AI\@.  We next present our experiments (Section~\ref{sec:experimental_setup}),  results (Section~\ref{sec:results}) and  further
discussion (Section~\ref{sec:discussion}). Finally, relevant
literature is discussed in Section~\ref{sec:related_works}, and
Section~\ref{sec:conclusions} summarizes our conclusions.


\section{The Brute-Force Dedispersion Algorithm}
\label{sec:dedispersion_kernel}
\label{sub:dispersion}

In dispersion \citep{lorimer2005}, the highest frequency
in a certain band
$f_{h}$ is received at time $t_{x}$, while lower simultaneously emitted frequency components $f_{i}$ arrive at $t_{x} + k$.
For frequencies expressed in MHz this delay in seconds is:
\begin{equation}
\label{eq:Dispersion}
k \approx 4150 \times DM \times \left(\frac{1}{f_{i}^{2}} - \frac{1}{f_{h}^{2}}\right)
\end{equation}
Here the Dispersion Measure \textit{DM} represents the
projected number of free electrons between the source and the receiver.
In incoherent dedispersion,  the lower frequencies
 are shifted in time and realigned  with the higher ones, thus
 approximating the original signal.

In a survey, the incoming signal must be brute-force dedispersed for
thousands of possible DM values.
As every telescope pointing direction or
\textit{beam} can be processed independently, performance of the
dedispersion algorithm can be improved by means of large-scale parallelization.


\subsection{Sequential Algorithm}
\label{sub:sequential_algorithm}

The input of this algorithm is a frequency-channelized time series,
represented as a $c \times t$ matrix,
with $c$ frequency channels and $t$ time samples
needed to dedisperse one second of data.
The output is a set of
 $d$ dedispersed trial-DM time-series, each of length $s$ samples per second, represented
as a $d \times s$ matrix.  All data are single precision floating point numbers; their real-life rates are e.g.
36~GB/s input and 72~GB/s output for the pulsar search with Apertif on the Westerbork telescope \citep{leeu14}.

Dedispersion (sequential pseudocode shown in Algorithm~\ref{alg:Dedispersion}) then consists of three nested loops, and
every output element is the sum of $c$ samples: one for each frequency
channel.  Which samples are part of each sum depends on the applied
delay (i.e. $\Delta$) that, as we know from
Equation~\ref{eq:Dispersion}, is a non-linear function of frequency
and DM\@.  These delays
can be computed in advance, so they do not contribute to the
algorithm's complexity.  Therefore, the complexity of this algorithm is $O(d \times s \times c)$.

\begin{algorithm}
\caption{Pseudocode of the brute-force dedispersion algorithm.}
\label{alg:Dedispersion}
\footnotesize{\begin{algorithmic}
\FOR{dm = 0 $\to$ d}
  \FOR{sample = 0 $\to$ s}
    \STATE{dSample = 0}
    \FOR{chan = 0 $\to$ c}
      \STATE{dSample += input[chan][sample + $\Delta$(chan, dm)]}
    \ENDFOR
    \STATE{output[dm][sample] = dSample}
  \ENDFOR
\ENDFOR
\end{algorithmic}}
\end{algorithm}

In the context of many-core accelerators, there is another, extremely important algorithmic
characteristic: the \textit{Arithmetic
  Intensity} (AI), i.e.\ the ratio between the performed
floating-point operations and the number of bytes accessed in global
memory.  In many-core
architectures the gap between computational capabilities and memory
bandwidth is wide, and a high AI is thus a prerequisite for high
performance \citep{williams2009}.  Unfortunately,
the AI for Algorithm~\ref{alg:Dedispersion} is
inherently low, with only one floating point operation per four bytes loaded from global memory.
For dedispersion,
\begin{equation}
\label{eq:DedispersionAI}
AI = \frac{1}{4 + \epsilon} < \frac{1}{4}
\end{equation}
where
$\epsilon$ represents the effect of accessing the delay table and
writing the output.
This low AI shows that brute-force dedispersion
is memory bound on most architectures. Its
performance is thus limited not by computational
capabilities, but by memory bandwidth.  One way to increase
AI and thus improve
performance, is to reduce the number of reads from global memory, by
implementing some form of data reuse.  In
Algorithm~\ref{alg:Dedispersion} some data reuse appears possible.
For some frequencies, the delay
is the same for two close DMs, $dm_{i}$ and $dm_{j}$, so that
$\Delta(c, dm_{i}) = \Delta(c, dm_{j})$.  Then, one input
element provides two different sums. With data reuse,
\begin{equation}
\label{eq:DedispersionAIp}
\footnotesize{AI < \frac{d \times s \times c}{4 \times ((s \times c) + (d \times s) + (d \times c))}  = \frac{1}{4 \times (\frac{1}{d} + \frac{1}{s} + \frac{1}{c})}}
\end{equation}
The bound from Equation~\ref{eq:DedispersionAIp} theoretically goes toward
infinity, but in
practice the non-linear delay function diverges rapidly at lower
frequencies. There is
never enough data reuse to approach the upper bound of
Equation~\ref{eq:DedispersionAIp}; for a more in-depth discussion see~\cite{sclocco2014}.
We thus categorize the algorithm as memory-bound.
In this conclusion we differ from previous literature such as
\cite{barsdell2010} and~\cite{barsdell2012b}.
The importance of the above mentioned data reuse in dedispersion was identified early on and implemented in e.g.\ the tree dedispersion algorithm \citep{tay74}.
That fast implementation has the drawback of assuming the dispersion sweep is linear.
Several modern pulsar and FRB surveys with large fractional bandwidths have used modified tree algorithms that sum over the \emph{quadratic} nature of the sweep (e.g.\ \cite{mlc+01}, \cite{masui2015}).



\subsection{Parallelization}
\label{sub:parallelization}

We first determine how to divide and organise the work of different
threads, and describe these in  OpenCL terminology.
We identify three main algorithm dimensions: DM, time and
frequency. Time and DM
are independent, and
ideal for parallelization, avoiding any inter- and
intra-thread dependency.  In our implementation, each OpenCL work-item
(i.e.\ thread) is associated with a different (DM, time) pair and it
executes the innermost loop of Algorithm~\ref{alg:Dedispersion}.  An
OpenCL work-group (i.e.\ group of threads) combines work-items that
are associated with the same DM, but with different time samples.

This proposed organization also simplifies memory access, using
coalesced reads and writes. Different small
requests can then be combined in one bigger operation. This well-known
optimization is a
performance requisite for many-core architectures, especially for
memory-bound algorithms like dedispersion.  Our output
elements are written to adjacent, and aligned, memory locations. The
\emph{reads} from global memory are also coalesced but, due to the shape of
the delay function, are not always aligned.
The worst-case memory overhead is at most a factor two \citep{sclocco2014}.

To exploit data reuse, we compute more than
one DM per work-group.  So, the final structure of our many-core
dedispersion algorithm consists of two-dimensional work-groups.  In
this way a work-group is associated with more than one DM, so that its
work-items can either collaborate to load the necessary elements from
global to \textit{local memory} (a fast memory area that is shared
between the work-items of a same work-group) or rely on the cache.

The general structure of the algorithm can be specifically
instantiated by configuring five user-controlled parameters.  Two
parameters control the number of work-items per work-group
in the time and DM dimensions, thus regulating
parallelism. Two parameters control the number of
elements per work-item. The last
parameter specifies whether local memory or
cache are employed for data reuse. These parameters determine source code generation at run time.
Because we do not know, a priori, the optimal configuration of these parameters, we rely on auto-tuning, i.e.\ we try all meaningful combinations and select the best performing one.



\section{Experimental Setup}
\label{sec:experimental_setup}

In this section we describe how the experiments are carried out; all
information necessary for replication is detailed in \citet{sclocco2014}.
Table~\ref{tab:Platforms} lists  the platforms  used, and reports
basic details such as number of OpenCL Compute Elements (CEs) (i.e.\ cores), peak performance, peak memory bandwidth and thermal design power (TDP).

\begin{table}
\centering
\footnotesize{\begin{tabular}{| l | r | r | r | r |}
\hline
\textbf{Platform} & \textbf{CEs} & \textbf{GFLOP} & \textbf{GB} & \textbf{Watt} \\
\textbf{} & \textbf{}    & \textbf{/s} & \textbf{/s} & \textbf{} \\
\hline
AMD HD7970 & $64 \times 32$ & 3,788 & 264 & 250 \\
\hline
NVIDIA GTX 680 & $192 \times 8$ & 3,090 & 192 & 195 \\
\hline
NVIDIA GTX Titan & $192 \times 14$ & 4,500 & 288 & 250 \\
\hline
NVIDIA K20 & $192 \times 13$ & 3,519 & 208 & 225 \\
\hline
Intel Xeon Phi 5110P & $2 \times 60$ & 2,022 & 320 & 225 \\
\hline
Intel Xeon E5-2620 & $6 \times 1$ & 192 & 42 & 95 \\
\hline
\end{tabular}}
\caption{Characteristics of the used platforms.}
\label{tab:Platforms}
\end{table}

We run the same code on every many-core accelerator; the source code\footnote{https://github.com/isazi/Dedispersion} is implemented in C++ and OpenCL\@.
The accelerators are part of the Distributed ASCI Supercomputer 4 (DAS-4)~\footnote{http://www.cs.vu.nl/das4/}.
As dedispersion is usually part of a larger pipeline,  input and output
are assumed to reside on the device. We thus do not measure data transfers over the PCI-e bus.

\begin{table}
\centering
\footnotesize{\begin{tabular}{| l | r | r | r | r |}
\hline
\textbf{Survey} & \textbf{s (Hz)}& \textbf{BW (MHz)} &  \textbf{n$_{chan}$} & \textbf{f (MHz)}  \\
\hline
LOFAR   & 200,000 & 6     & 32 & 138$-$145 \\
\hline
Apertif & 20,000  & 300 & 1,024 & 1,420$-$1,729 \\
\hline
\end{tabular}}
\caption{Survey name, sampling rate $s$, bandwidth $BW$, total number of
  channels n$_{chan}$ and frequency range $f$ for the two experimental setups.}
\label{tab:Surveys}
\end{table}

The experimental set ups (Table~\ref{tab:Surveys}) are based on two different pulsar surveys,
one for a hypothetical high-time resolution LOFAR survey comparable to the LOFAR Pilot Pulsar
Survey \citep{clh+14} and one for the planned Apertif pulsar/FRB survey \citep{leeu14}.
For both, trial DMs start at 0 and increment by  0.25 $pc/cm^{3}$.
These two setups stress different characteristics of the algorithm --
the Apertif setup, at 20~MFLOP per DM, is three times more compute
intensive than LOFAR at 6~MFLOP per DM\@.
Yet the higher Apertif frequencies cause reduced  delays, with more potential for data reuse.
We thus try two complementary scenarios: one is more computationally intensive, but potentially offers more data reuse, and one that is less computationally intensive, but precludes almost any data reuse.

We auto-tune the five algorithm parameters described in
Section~\ref{sec:dedispersion_kernel}, for each of the six platforms of Table~\ref{tab:Platforms}, in both observational setups.
We use 12 different input instances between 2$-$4,096
\citep{sclocco2014}, and measure our performance metric, the number of single precision floating-point operations per second.


\section{Results}
\label{sec:results}

\subsection{Auto-Tuning}
\label{sub:auto_tuning}

For the Apertif case,
 the optimal number of work-items per work-group identified by
auto-tuning is the highest for the GTX 680 (1,024). The other three GPUs require between 256 and 512, while the two Intel platforms
require the lowest number (i.e.\ between 16 and 128). As detailed in
\citet{sclocco2014}, the optimal configuration is more variable with
smaller input instances, where optimization
space is small.

Results for the LOFAR setup are more stable. It has less available
data reuse, easing the identification of the optimum. The number of  work-items per work-group
stay similar for the GPUs, and is somewhat lowered for the Intel platforms.

Even for similar total numbers of work-items per work-group, the two
underlying parameters (Section~\ref{sec:dedispersion_kernel}) may
differ.  For e.g.\ the HD7970, the Apertif work-group is as $8 \times
32$ work-items, while LOFAR it is composed by $64 \times 4$
work-items.  In the Apertif setup the auto-tuning identifies a
configuration that intensively exploits the available data reuse,
while in the LOFAR setup the optimal configuration relies more on the
device occupancy.

This result is important:
accelerated dedispersion algorithm has no single
optimal configuration -- it must be adapted to the exact observational
setup.

The subsequent auto tuning the amount of work per work-item
again exploits each accelerator's advantage, such as the high number of registers in the K20 and
Titan, adapting the algorithm to different scenarios \citep{sclocco2014}.

The last tunable parameter is the explicit use of local memory,
present on the GPUs, to exploit data reuse, over just hardware
cache. Again auto-tuning adjusted the interaction of different parameters
for highest performance dedispersion in each platform.

In summary, we find that optimal configurations cannot be identified a
priori, and that auto-tuning is the only feasible way to properly
configure the dedispersion algorithm, because of the number and
interaction of parameters, and their impact on AI\@.


\subsection{Impact of Auto-Tuning on Performance}
\label{sub:impact_of_auto_tuning}

\begin{figure}
\centering
\includegraphics[width=\columnwidth]{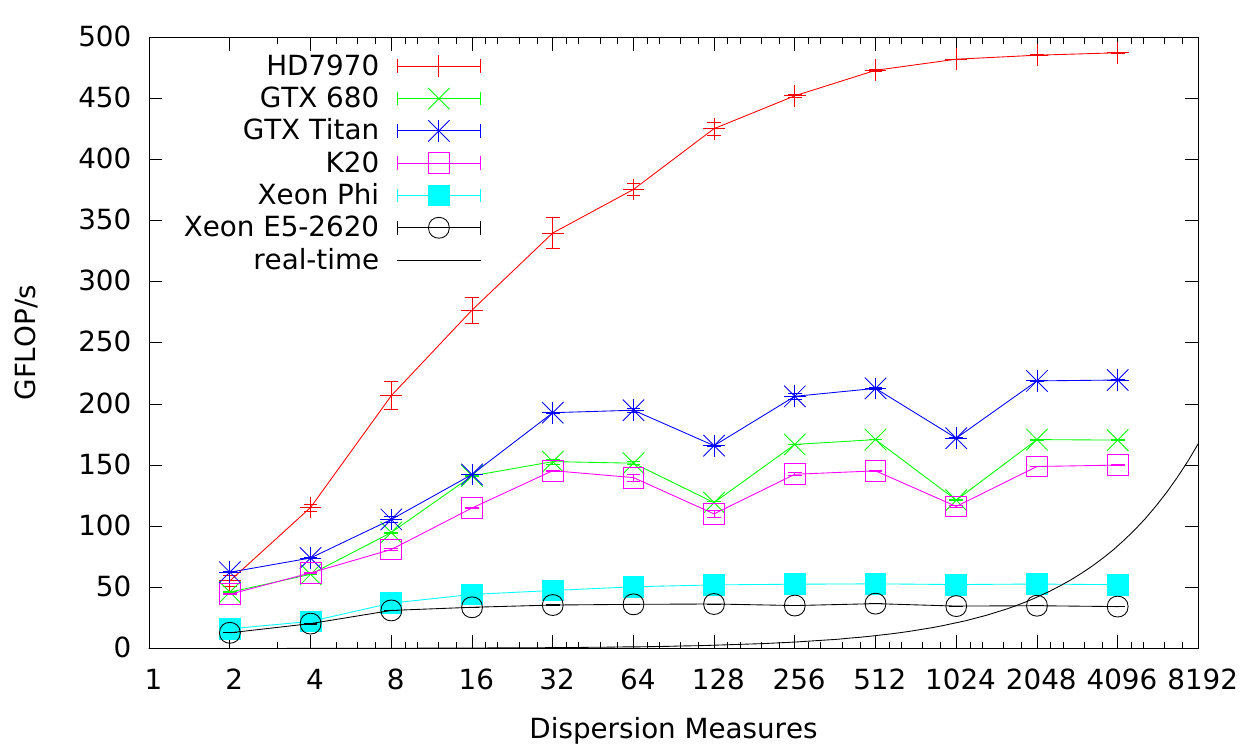}
\caption{Performance of auto-tuned dedispersion, Apertif (higher is better).}
\label{fig:PerfApertif}
\end{figure}

In Fig.~\ref{fig:PerfApertif} we show the performance achieved by
auto-tuned dedispersion for the Apertif case.
All platforms show a performance increase with the dimension of the
input instance, and plateauing afterwards.
Note how the tuned algorithm scales better than linearly up to this maximum, and then scales linearly.
The HD7970 achieves highest performance, the Intel CPU and the Xeon
Phi are at the bottom, and the three NVIDIA GPUs occupy the middle of this figure.
On average, the HD7970 is 2.7 times faster than the NVIDIA GPUs, and 11.3 times faster than the Intel CPU and Xeon Phi; the Phi is 1.5 times faster than the multi-core CPU.

\begin{figure}
\centering
\includegraphics[width=\columnwidth]{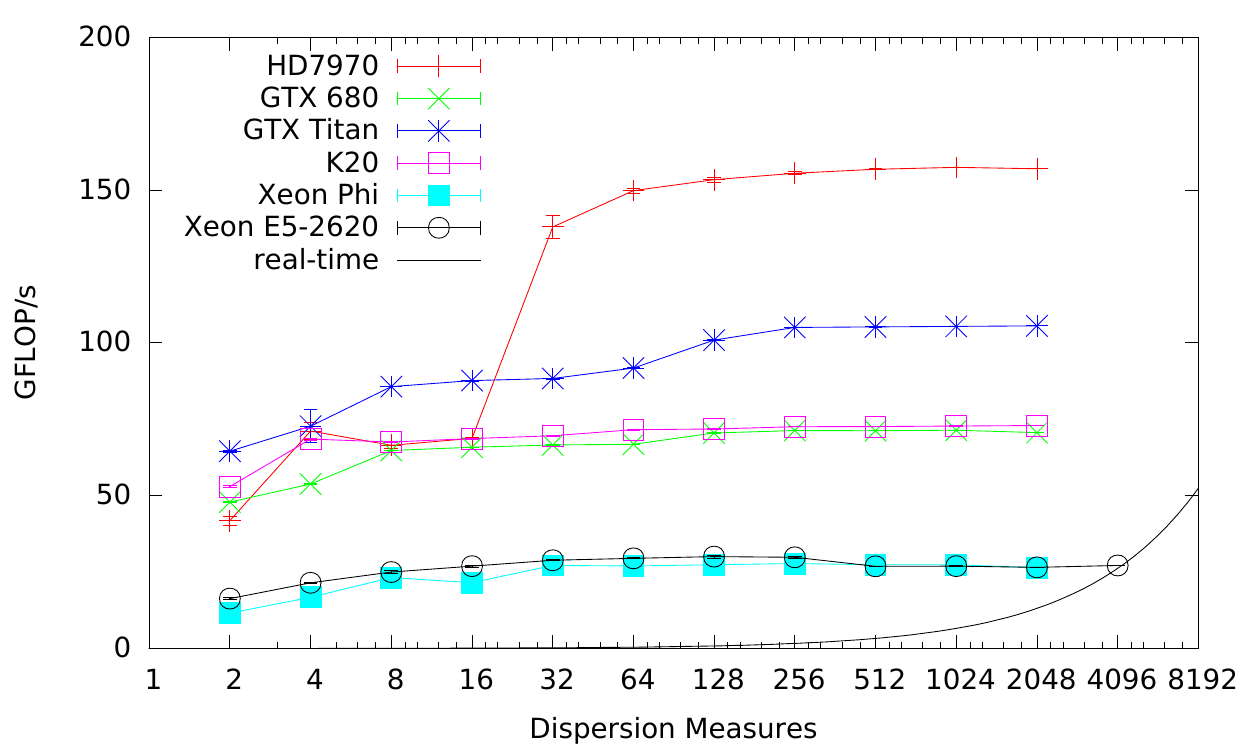}
\caption{Performance of auto-tuned dedispersion, LOFAR (higher is better).}
\label{fig:PerfLOFAR}
\end{figure}

The results for the LOFAR scenario (Fig.~\ref{fig:PerfLOFAR}) are different.
Absolute performance is lower.
This is caused by the reduced available data reuse, resulting in lower
algorithm AI\@.
Next, the GPUs are closer in performance; now the HD7970 is only 1.4
times faster than the GTX Titan.
With less data reuse available, memory bandwidth becomes increasingly
more important, leveling the differences. On average, the HD7970 is just 1.9 times faster than the NVIDIA GPUs, and 6 times faster than the Intel CPU and Xeon Phi.

In both figures, only scenarios above the ``\textit{real-time}'' line can dedisperse one second of data in less than one second of computation.
This is a fundamental requirement for modern radio telescopes, whose
 extreme data rate precludes data storage and off-line
processing. We show that GPUs can scale to bigger instances and/or to a multi-beam scenario, and still satisfy the real-time constraints.

\begin{figure}
\centering
\includegraphics[width=\columnwidth]{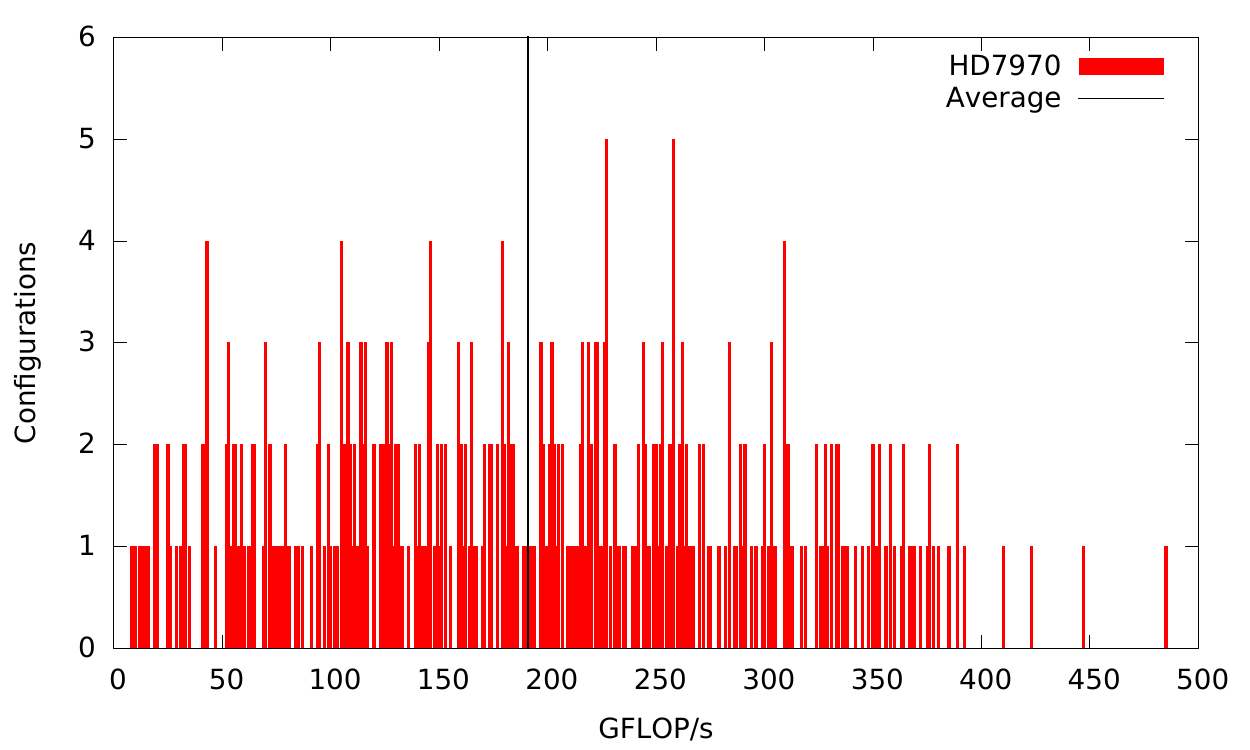}
\caption{Example of a performance histogram, for the case or Apertif 2,048 DMs.}
\label{fig:HD7970HistApertif}
\end{figure}

What was the impact of auto-tuning on performance?
Figure~\ref{fig:HD7970HistApertif} shows the histogram of performance in the optimization space.
The optimum lies far from the typical configuration.
Because optimal configurations depends on multiple parameters, like the platform used to execute the algorithm, and specific observational parameters, it will not be trivial, even for an experienced user, to manually select the best configuration without extensive tuning.
This claim is also supported by the high signal-to-noise ratio of the optimal configurations, as detailed in \citet{sclocco2014}.

In summary, we satisfy a realistic real-time
constraint in almost every test case, which allows for massive
multi-beaming.
Optimal configurations are difficult to guess in this optimization space, and therefore auto-tuning has a critical impact on the performance.


\subsection{Data Reuse and Performance Limits}
\label{sub:data_reuse_and_performance_limits}

To simulate a scenario with perfect data reuse, we tune and measure the performance of dedispersion using the value zero for all DM trials.
In the case of
For Apertif, the difference is negligible (cf.~Fig.~11 in
\citealt{sclocco2014}). Data reuse was already maximized.
For the LOFAR setup (cf.~Fig.~12 in \citealt{sclocco2014}), the performance improves, approaching the Apertif setup.
The increased data reuse is exploited until the hardware is saturated.

Performance is predominantly determined by the amount of possible data reuse, which
is a function of real-life frequencies and DM values.
This tests our hypothesis that even
with perfect data reuse,  high AI cannot be  achieved  because of the
limitations of real hardware (in contrast with the conclusions of~\citealt{barsdell2010}).
We therefore conclude that brute-force dedispersion is memory-bound for every practical and realistic scenario.


\section{Discussion}
\label{sec:discussion}

We first compare the performance of the \emph{auto-tuned} versus a \textit{generically tuned}  algorithm.
In the latter, tuning is confined to a mono-dimensional configuration of the work-groups and the work-items compute only one DM value.
This strategy is widely applied by programmers; but as the
algorithm's AI is unaffected, data reuse is not optimized.

\begin{figure}
\centering
\includegraphics[width=\columnwidth]{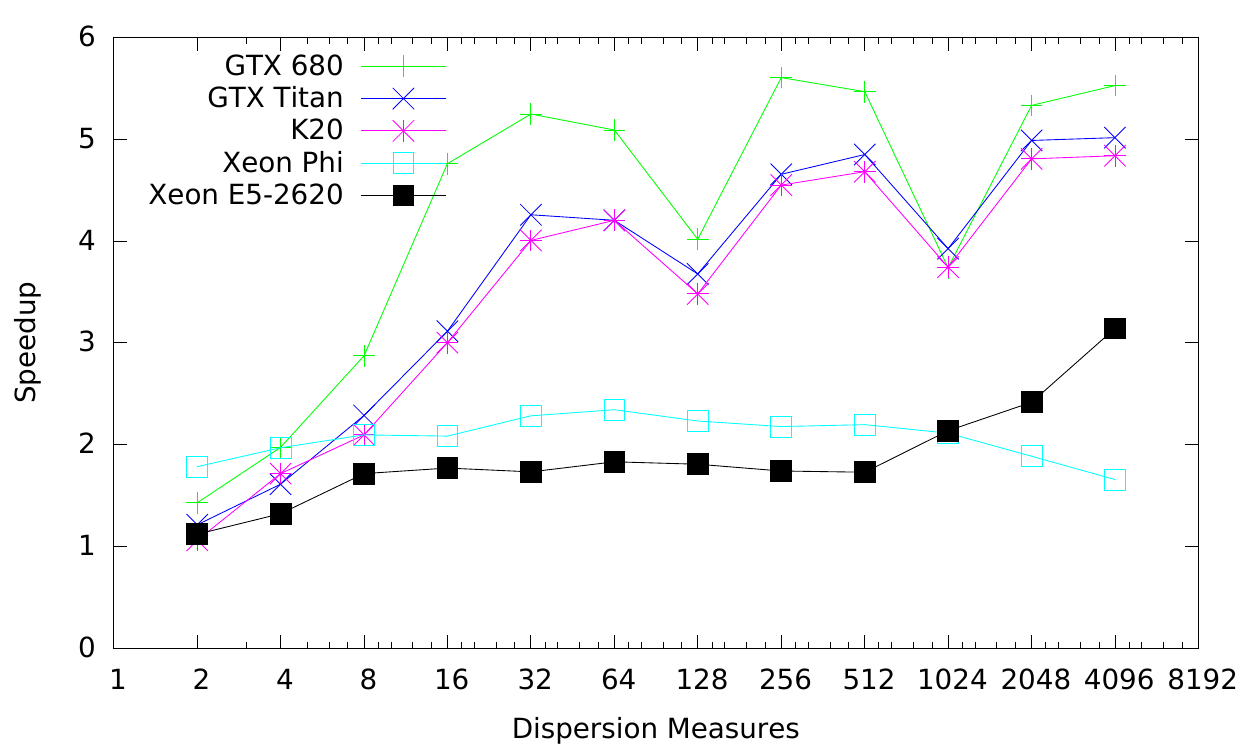}
\caption{Speedup over a generically tuned configuration, Apertif
  (higher is better). The HD7970 GPU is missing because
  there are no valid configurations of the
  algorithm that do not exploit data reuse.
}
\label{fig:SpeedupNoReuseApertif}
\end{figure}

We find that \emph{auto-tuned} dedispersion on GPUs is $\sim$5 times faster than
\emph{generically tuned} in the Apertif setup (Fig.~\ref{fig:SpeedupNoReuseApertif}), and still 1.5--2.5 times faster in the LOFAR setup.
Effectively exploiting data reuse is a main performance driver for a high-performance dedispersion algorithm, especially for those platforms where the gap between floating point peak performance and memory bandwidth is wide.

\begin{figure}
\centering
\includegraphics[width=\columnwidth]{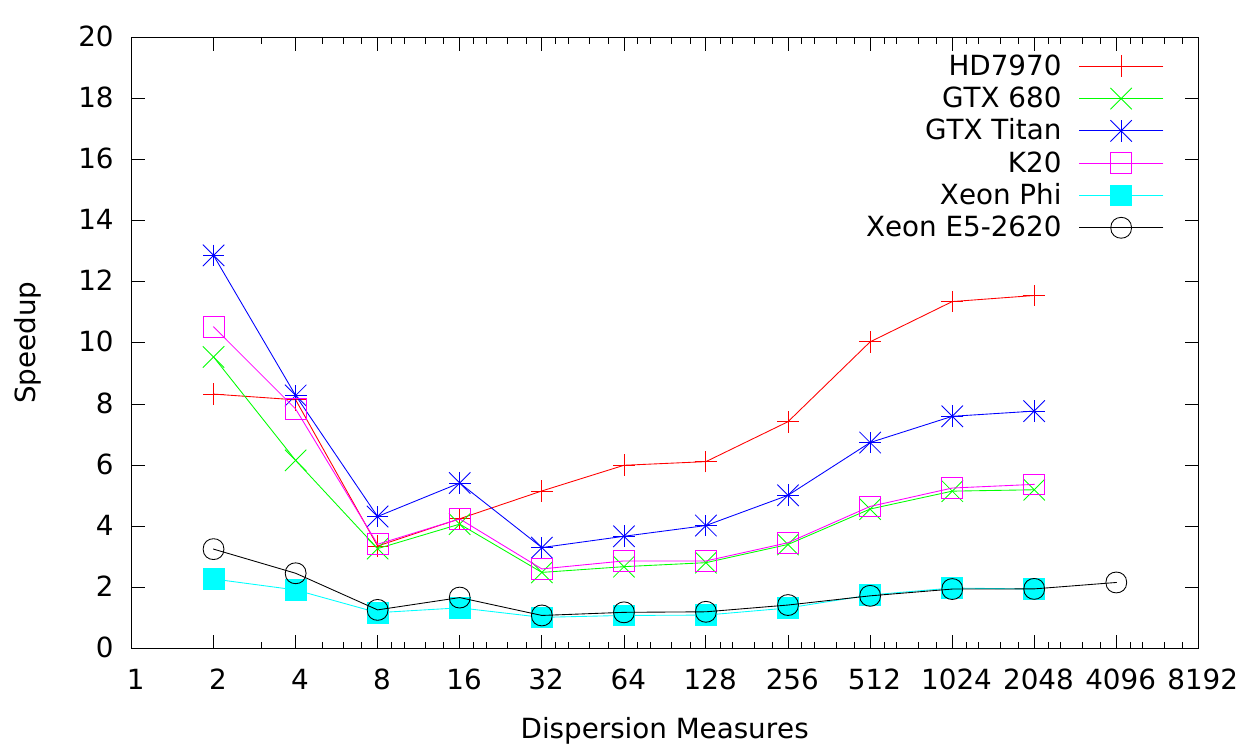}
\caption{Speedup over a parallel CPU implementation, LOFAR (higher is better).}
\label{fig:SpeedupLOFAR}
\end{figure}

We also compare the performance of our auto-tuned algorithm to an optimized CPU version.
This CPU version of the algorithm is parallelized using OpenMP instead of OpenCL, with different threads computing different chunks of DM values and time samples.
It includes all meaningful optimizations described in Section~\ref{sub:parallelization}, and is vectorized using Intel's Advanced Vector Exensions (AVX) intrinsics.
Like the OpenCL version, it is automatically auto-tuned.
For both observational setups (cf.~Fig.~\ref{fig:SpeedupLOFAR} for LOFAR), the many-core implementation is considerably faster than a highly optimized, and tuned, CPU implementation.

\begin{figure}
\centering
\includegraphics[width=\columnwidth]{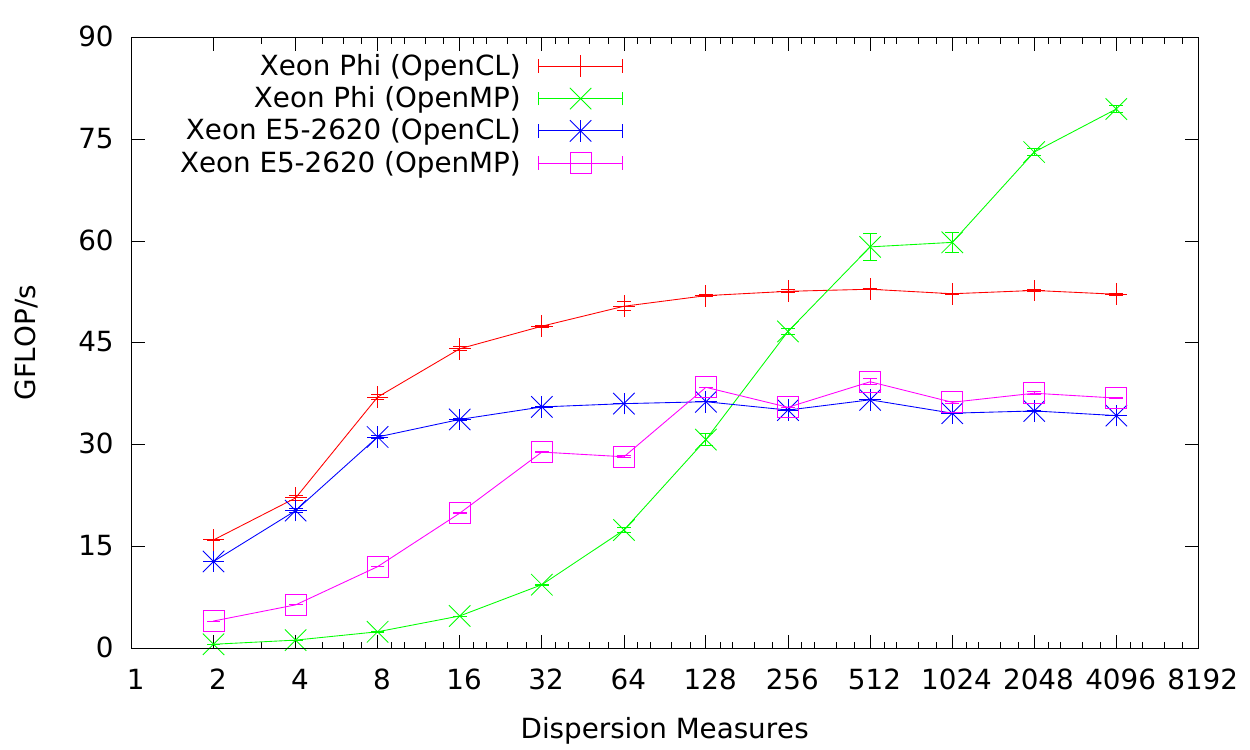}
\caption{Comparing OpenCL and OpenMP implementations on the CPU and Xeon Phi, Apertif (higher is better).}
\label{fig:XeonComparisonApertif}
\end{figure}

The OpenMP code generator was also extended to generate native code for the Xeon Phi, and we compared the Intel CPU and Xeon Phi, using both OpenCL and OpenMP\@.
In the Apertif case (Figure~\ref{fig:XeonComparisonApertif}) the OpenCL implementation is faster than the OpenMP one for smaller inputs, while the opposite is true for larger ones, and this behavior holds for both platforms.
In the LOFAR case, however, this same behavior applies only to the Xeon Phi, with the Xeon E5-2620 showing a performance degradation for bigger inputs that is consistent with the experiments of~\cite{armour2012}.
Overall improvements to the Intel OpenCL compiler could improve the Xeon Phi's performance, as we were able to achieve a 51\% improvement in performance on this platform using a tuned OpenMP implementation.

\subsection{Multi-Beam Performance}

In the conclusions of Section~\ref{sub:impact_of_auto_tuning} we highlighted how dedispersion performance can allow for  multi-beam scenarios.
Apertif will be need to dedisperse 2,000 DMs in real-time for 450 different beams, thus requiring 18.5 TFLOP/s just for dedispersion.
Using our best performing accelerator, the AMD HD7970, it is possible to dedisperse 2,000 DMs in 0.08 seconds; 9 beams per GPU could be dedispersed in real-time. Enough memory is available.
Therefore, the 18.5 TFLOP/s dedispersion instrument for Apertif could be implemented using 2015 technology with $\sim$50 GPUs.
With a HD7970 model with double the memory,  12 beams could be dedispersed at once, reducing the system size to only 38 GPUs.

For its real-life impact, we compute an upper bound on the power necessary for dedispersion on Apertif from the TDPs (Table~\ref{tab:Platforms}): the proposed GPU solution requires 12.5~kW while the traditional CPU solution requires 46.5~kW.
We can therefore conclude that, in this scenario, using many-core accelerators does not only provide a 9.8 times reduction in the size of the system, when compared to a traditional CPU-based solution, but it also lowers the power consumption by a factor of 3.7.
This improvement in energy efficiency is in part architectural, and in part caused by auto-tuning.
Using the results of Section~\ref{sub:impact_of_auto_tuning} it is possible to determine that the power that would be required by an average configuration is 22.5 kW:\@ we can thus conclude that auto-tuning is responsible for around 50\% of the total power savings.

\section{Related Work}
\label{sec:related_works}

In the literature, auto-tuning is considered a viable technique to achieve performance that is both high and portable.
In particular, \cite{li2009} show that it is possible to use auto-tuning to improve performance of even highly-tuned algorithms, and \cite{datta2008} affirm that application specific auto-tuning is the most practical way to achieve high performance on multi-core systems.
Highly relevant here is the work of~\cite{du2012}, with whom we  agree  that auto-tuning can be used as a performance portability tool, especially with OpenCL\@.
Another attempt at achieving performance portability on heterogeneous systems has been made by~\cite{phothilimthana2013}, and while their approach focuses on the compiler, it still relies on auto-tuning to map algorithms and heterogeneous platforms in an optimal way.
In recent years, we have been working on parallelizing and implementing radio astronomy kernels on multi and many-core platforms \citep{nieuwpoort2011}, and using auto-tuning to achieve high performance on applications like the LOFAR beam former \citep{sclocco2012}.
What makes our current work different, is that we do not only use auto-tuning to achieve high performance, but measure its impact, and show that the optimal configurations are difficult to guess without thorough tuning.

While there are no previous attempts at auto-tuning dedispersion for many cores, there are a few previous GPU implementations documented in literature.
First, in \cite{barsdell2010} dedispersion is listed as a possible candidate for acceleration, together with other astronomy algorithms.
While we agree that dedispersion is a potentially good candidate for many-core acceleration because of its inherently parallel structure, we believe their performance analysis to be too optimistic, and the AI estimate in \citet{barsdell2010} to be unrealistically high.
In fact, we showed in this paper that dedispersion's AI is low in all realistic scenarios, and that the algorithm is inherently memory-bound.
The same authors implemented, in a follow-up paper~\citep{barsdell2012b}, dedispersion for NVIDIA GPUs, using CUDA as their implementation framework.
However, we do not completely agree with the performance results presented in \citet{barsdell2012b} for two reasons: first, they do not completely exploit data reuse, and we have shown here how important data reuse is for performance; and second, part of their results are not experimental, but derived from performance models.

Another GPU implementation of the dedispersion algorithm is presented in~\cite{magro2011}.
Also in this case there is no mention of exploiting data reuse.
In fact, some of the authors of this paper published, shortly after~\cite{magro2011}, another short paper \citep{armour2012} in which they affirm that their previously developed algorithm does not perform well enough because it does not exploit data reuse.
Unfortunately, this paper does not provide sufficient detail on either the algorithm or on experimental details, such as frequencies and time resolution, for us to repeat their experiment.
Therefore, we cannot verify the claimed 50\% of theoretical peak performance.
However, we believe this claim to be unrealistic because dedispersion has an inherently low AI, and it cannot take advantage of fused multiply-adds, which by itself already limits the theoretical upper bound to 50\% of the peak on the used GPUs.

A different approach to both brute-force and tree based dedispersion has been proposed by \citet{zackay2014}.
This new algorithm has lower computational complexity than brute-force dedispersion, and appears to be faster than the implementations of both \citet{magro2011} and \citet{barsdell2012b}.
Although the experimental results highlighted in our paper show higher performance than the results presented in \citet{zackay2014}, we believe this algorithm has potential; and that it, too, could benefit from auto-tuning to further improve performance in real-life implementations.


\section{Conclusions}
\label{sec:conclusions}

In this paper, we analyzed the brute-force dedispersion algorithm, and presented  our many-core implementation. We analytically proved that dedispersion is a memory-bound algorithm and that, in any real world scenario, its performance is limited by low arithmetic intensity.
With the experiments presented in this paper, we also demonstrated that by using auto-tuning it is possible to obtain high performance for dedispersion.
Even more important, we showed that auto-tuning makes the algorithm portable between different platforms and different observational setups.
Furthermore, we highlighted how auto-tuning permits to automatically exploit the architectural specificities of different platforms.

Measuring the performance of the tuned algorithm, we verified that it scales linearly with the number of DMs for every tested platform and observational setup.
So far, the most suitable platform to run dedispersion among the ones we tested, is a GPU from AMD, the HD7970.
This GPU performs 2--9 times better than the other accelerators when extensive data reuse is available, and remains 1.4--6 times faster even in less optimal scenarios.
Overall, the GPUs are 4.9--7.5 and 3.8 times faster than the CPU and Xeon Phi in the two scenarios analyzed in this work.
We conclude that, at the moment, GPUs are the best candidate for brute-force dedispersion.

In this paper, we showed that using the HD7970 GPU it would be possible to
build the 18.5 TFLOP/s dedispersion instrument for Apertif using just $\sim$50 accelerators, thus reducing the system size of a factor 9.8 and the power requirements of a factor 3.7.
These improvements are only in part architectural, and mostly a result of optimal tuning of the algorithm.
We will continue to compare platforms as the design date for SKA approaches.

Another important contribution of this paper is the quantitative evidence of the impact that auto-tuning has on performance.
With our experiments, we showed that the optimal configuration is difficult to find manually and lies far from the average.
Moreover, we showed that the auto-tuned algorithm is faster than a generically tuned version of the same algorithm on all platforms, and is an order of magnitude faster than an optimized CPU implementation.

Finally, our last contribution was to provide further empirical proof that brute-force dedispersion is a memory-bound algorithm, limited by low AI\@.
In particular, we showed that achievable performance is limited by the amount of data reuse that dedispersion can exploit, and the available data reuse is affected by parameters like the DM space and the frequency interval.
We also showed that, even in a perfect scenario where data reuse is unrealistically high, the performance of dedispersion is limited by the constraints imposed by real hardware, and approaching the theoretical AI bound of the algorithm becomes impossible.


\bibliographystyle{elsarticle-num-names}
\bibliography{Bibliography}

\end{document}